\documentclass[a4paper]{jpconf}
\usepackage{graphicx}

\def\xb{\overline{x}}

\begin{document}
\title{Role of pion pole in hard exlusive meson leptoproduction}

\author{Sergey Goloskokov}

\address{Bogoliubov Laboratory of Theoretical
Physics, Joint Institute for Nuclear Research, Dubna 141980,
Moscow region, Russia}

\ead{goloskkv@theor.jinr.ru}

\begin{abstract}
We consider the pion pole contribution and transversity effects
determined by the  $H_T$ and $\bar E_T$ Generalized Parton
Distributions (GPDs) which are essential in hard pseudoscalar and
vector meson leptoproduction. We investigate spin effects in the
$\omega$ and $\rho^0$ reactions. It is shown that the pion pole
contribution is very important in the $\omega$ production. Such
effects in the $\rho^0$ channel are much smaller. Our results on
spin asymmetries and spin density matrix elements in these
reactions were found to be in good agreement with HERMES
data.\end{abstract}

\section{Introduction}
In this report, we  discuss pion pole effects \cite{gk14} and
transversity twist-3 contributions from the  $H_T$, $\bar E_T$
GPDs \cite{gk09} which are important in explanation of spin
effects in the light meson production at moderate $Q^2$. Our model
of hard electroproduction of light mesons \cite{gk06} is based on
the handbag factorization at high photon virtuality $Q^2$ of the
process amplitude  into the hard meson electroproduction off
partons, and GPDs \cite{fact}.

The amplitudes of the pseudoscalar meson (PM) leptoproduction  in
the leading twist approximation are sensitive to GPDs
$\widetilde{H}$ and $\widetilde{E}$.   These contributions  were
found to be not sufficient to describe spin effects in the PM
production at sufficiently low $Q^2$ \cite{gk09}. To be consistent
with experiment, one needs the essential contributions from the
transversity GPDs $H_T$, $\bar E_T$ which go together  with a
twist-3 pion wave function \cite{gk11}. We discuss the pion pole
and transversity effects in the PM leptoproduction at HERMES and
CLAS energies in section 2. Our results are in good agreement with
experimental data.

The  HERMES data on  SDMEs for the $\omega$ production indicated
essential contributions from unnatural parity exchanges
\cite{omega14}. It was found that the pion pole (PP) contribution
\cite{gk14} is significant in explanation of the large
unnatural-parity effects observed by HERMES. The PP contribution
to the $\rho^0$ production is much smaller with respect to the
$\omega$ case. We discuss the PP effects in $\omega$ and $\rho$
production in section 3.

\section{Handbag approach. Pseudoscalar meson leptoproduction.}
Within the handbag approach the leading contributions to the meson
production amplitude off  proton  at sufficiently high photon
virtuality $Q^2$ can be described in factorized form \cite{fact}
as a convolution of a hard meson subprocess amplitude off partons
with the same helicities ${\cal H}^a_{\mu' +,\mu +}$ and GPDs as
\begin{equation}\label{ff}
M_{\mu' +,\mu +} \propto \int_{-1}^1 dx
   {\cal H}^a_{\mu' +,\mu +} F^a(x,\xi,t);\;\; M_{\mu' -,\mu +} \propto \frac{\sqrt{-t}}{2 m}
                          \int_{-1}^1 d\xb\,{\cal H}^a_{\mu' +,\mu
                          +}\,
           E^a(\xb,\xi,t).
\end{equation}
Here $a$ is a flavor factor and  $\mu$ and $\mu'$ are helicities
of the photon and produced meson.

GPDs  contain information on the hadron structure. With the help
of sum rules they are connected with hadron form factors, and
information on the parton angular momenta can be extracted. In the
forward limit $t=0$ and zero skewness $\xi=0$ GPDs are equivalent
to ordinary Parton Distribution Functions (PDFs). We estimate GPDs
by using the double distribution representation \cite{mus99} which
connects GPDs with PDFs. The PDFs parameterizations are obtained,
e.g, from the  analysis \cite{CTEQ6}, or from the nucleon form
factor study \cite{pauli}.

The handbag approach was successfully applied to light meson
leptoproduction \cite{gk06}. The cross sections and spin
observables of light vector meson (VM) leptoproduction were found
to be in good agreement with HERMES, COMPASS and HERA data.

The leading twist contributions are not sufficient to describe the
experimental results on PM electroproduction  at low $Q^2$. This
can be found from analysis of $A_{UT}^{\sin(\phi_s)}$ asymmetry
which in the leading twist handbag approximation is small.
However, HERMES found \cite{airappi} that this asymmetry is large,
about 0.5. This effect can be explained by the large contributions
to the $A_{UT}^{\sin(\phi_s)}$ asymmetry from the amplitude
$M_{0-,++}$. At low $Q^2$ the amplitudes $M_{0\pm,++}$ are
determined by  the transversity GPDs $H_T$ and $\bar E_T$
contributions which have the twist-3 nature. Within the handbag
approach the transversity GPDs are accompanied by a twist-3 meson
wave function in the hard subprocess amplitude ${\cal H}$
\cite{gk11} which is the same for both the $M^{tw-3}_{0\pm,++}$
amplitudes
\begin{equation}\label{ht}
M^{tw-3}_{0-,++} \propto \,
                            \int_{-1}^1 d\xb
   {\cal H}_{0-,++}(\xb,...)\,H_T;\;
   M^{tw-3}_{0+,++} \propto \, \frac{\sqrt{-t'}}{4 m}\,
                            \int_{-1}^1 d\xb
 {\cal H}_{0-,++}(\xb,...)\; \bar E_T.
\end{equation}

The $H_T$ GPDs in the forward limit and $\xi=0$ are equal to
transversity PDFs $\delta$ and are parameterized   by using the
model \cite{ans}. The double distribution  is used to calculate
GPDs as before. For details, see \cite{gk11}.

Information on $\bar E_T$ is obtained now only from the lattice
QCD \cite{lat}. The lower moments of $\bar E_T^u$ and $\bar E_T^d$
were found to be quite large, have the same sign and a similar
size.   At the same time  $H_T^u$  and $H_T^d$ GPDs have a
different sign. These properties of GPDs provide an essential
compensation of the $\bar E_T$ contribution in the $\pi^+$
amplitude, but $H_T$ effects are not small there. For the $\pi^0$
production we have large  $\bar E_T$ contributions and smaller
$H_T$ effects.

We present here our results on the PM  leptoproduction based on
the handbag approach. In calculation, we use the leading
contributions together with the transversity effects (\ref{ht})
which are essential at low $Q^2$.

In Fig.1, we show our results for the $\pi^+$ production which are
in agreement with HERMES \cite{pipl}. The dashed line indicates
the model results for $H_T=0$. It is closed to the PP contribution
to the $\pi^+$ cross section. So in this channel the PP effects
are very essential. The pion induced Drell-Yan process $\pi^- p
\to \gamma^* n \to l^+l^-n$ which is a time-like analog of $\pi^+$
leptoproduction was investigated in \cite{gkdy15}. It was found
that the pion pole gives the predominant contribution in the
longitudinal cross section of this process. Other effects are
rather small in $\sigma_L$.

 In Fig. 2, we present the model results for the $\pi^0$
 production cross section \cite{gk11}. Here the PP contribution is
 absent but transversity effects are essential.
  At small momentum transfer the $H_T$ contribution is
visible and provides a nonzero cross section. At $-t' \sim 0.2
\mbox{GeV}^2$ the $\bar E_T$ contribution becomes predominant and
gives a maximum in the cross section. A similar contribution from
 $\bar E_T$ is observed in the interference cross section
$\sigma_{TT}$. Note that the $\bar E_T$ effects in the unseparated
cross section $\sigma$ which is saturated by $\sigma_T$
 is strongly
correlated with similar effects  in $\sigma_{TT}$. The fact that
we describe well both the unseparated $\sigma$ and $\sigma_{TT}$
cross sections can indicate that transversity effects were
probably observed in CLAS \cite{bedl}.

\begin{figure}[h]
\begin{minipage}{17pc}
\includegraphics[width=17pc,height=6.1cm]{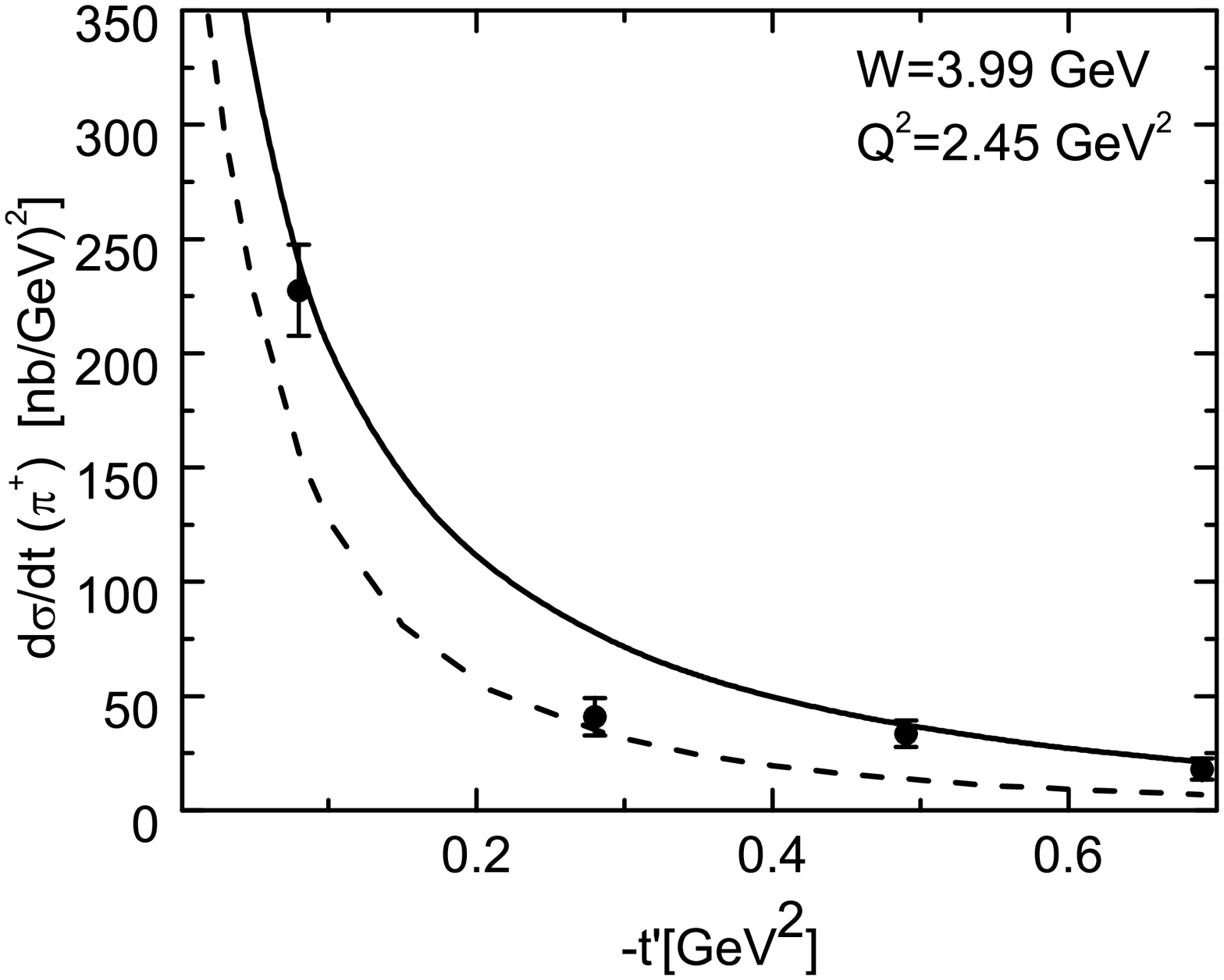}
\caption{\label{label2} Cross section of $\pi^+$ production at
HERMES energies with the data \cite{pipl}. Full line- unseparated
cross section. Dashed line- model results for $H_T=0$.}
\end{minipage}\hspace{2pc}%
\begin{minipage}{17pc}
\includegraphics[width=17pc,height=5.9cm]{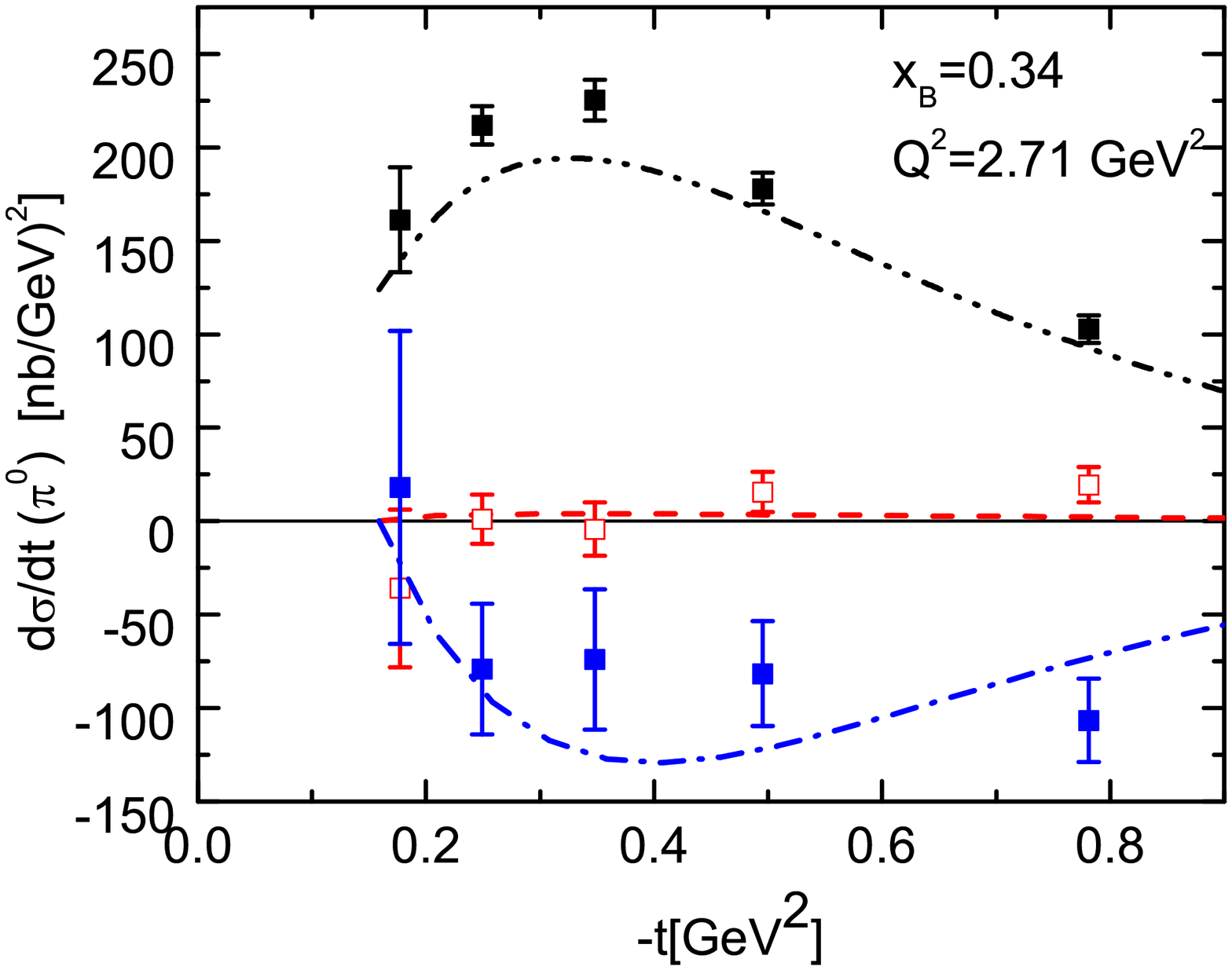}
\caption{\label{label2} $\pi^0$ production in the CLAS energy
range together with the data \cite{bedl}. Dashed-dot-dotted line-
$\sigma=\sigma_{T}+\epsilon \sigma_{L}$, dashed
line-$\sigma_{LT}$, dashed-dotted line- $\sigma_{TT}$.}
\end{minipage}
\end{figure}

\section{Large unnatural parity effects in $\omega$ production. Pion pole contribution.}

In most reactions the unnatural parity (UP) contributions are
small with respect to the natural ones. The  HERMES data on the
spin density matrix elements (SDMEs) for the $\omega$ production
indicate the strong contributions from UP effects \cite{omega14}.
 It was found that the ratio of the unnatural and  natural parity
cross section, which was expected to be small, is larger than
unity. This can be caused by the large pion pole contribution to
this process.

\begin{figure}[h]
\includegraphics[width=17pc]{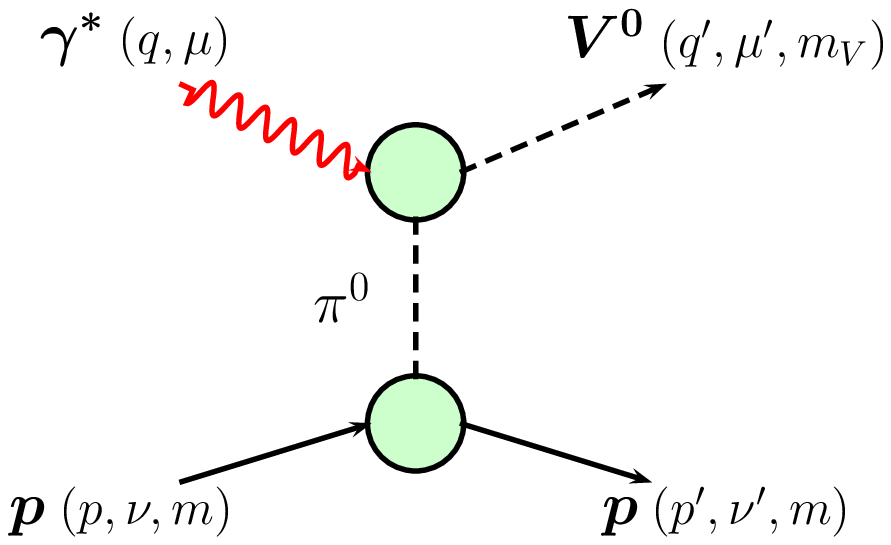}\hspace{2pc}%
\begin{minipage}[b]{17pc}\caption{\label{label} Pion pole contribution to VM
production. The UP helicity amplitudes determined by pion pole looks as
follows:\\[2mm]
{\large $M^{pole}_{++,++} \sim \frac{\rho_{\pi V}}{t-m^2_\pi}\,
\frac{m\,\xi\, Q^2}{\sqrt{1-\xi^2}}$,\;\\
 $M^{pole}_{+-,++} \sim -\frac{\rho_{\pi
V}}{t-m^2_\pi}\,\frac{\sqrt{-t'}\,Q^2}{2}$};\\with\\ $\rho_{\pi V}
\sim g_{\pi V}(Q^2)\, g_{\pi N\,N}\,F_{\pi N\,N}(t).$}
\end{minipage}
\end{figure}

In Fig.3, we show the PP contribution to the VM production
together with the helicity amplitudes generated by the pion pole.
These amplitudes have the UP nature and are controlled by the $\pi
V$ transition form factor.  The transition form factor $g_{\pi
V}(0)$ can be determined from the VM radiative decay
\begin{equation}
\Gamma(V \to \pi \gamma) \sim \frac{\alpha_{elm}}{24}\, |g_{\pi
V}(0)|^2M_V^3.
\end{equation}
We find
\begin{equation}
|g_{\pi \omega}(0)|=2.3 \mbox{GeV}^{-1}; \;\;\;\;\;\;|g_{\pi
\rho}(0)|=.85\mbox{GeV}^{-1}.
\end{equation}
 This means that $|g_{\pi \omega}(0)|$ is about 3 times
larger with respect to $|g_{\pi \rho}(0)|$ and we should observe
large PP effects in $\omega$ and small in $\rho$ production. The
$Q^2$ dependence of $g_{\pi V}(Q^2)$ was extracted \cite{gk14}
from the HERMES data \cite{omega14} on the ratio of the unnatural
to the natural parity cross section $U_1$  at $Q^2 <
4\mbox{GeV}^2$.

  We will discuss our results on PP effects
in the $\omega$ and $\rho$ production \cite{gk14} and give a
comparison  with  HERMES data \cite{omega14}. In calculations we
use GPDs from our analysis of hard meson leptoproduction.

The natural and unnatural parity asymmetry $P$ is determined as
follows:
\begin{equation}
P=\frac{d \sigma^N (\gamma^*_T \to V_T)-d \sigma^U (\gamma^*_T \to
V_T)}{d \sigma^N (\gamma^*_T \to V_T)+d \sigma^U (\gamma^*_T \to
V_T)}.
\end{equation}
If the UP contribution is small, we find $P \sim 1$. If it is
large,  the value of the $P$ asymmetry will be far from unity. The
model results for this asymmetry for $\omega$ production are shown
in Fig.4.  We find that  the PP contribution to the $\omega$
asymmetry give $P \sim -0.5$ (full line) in agreement with
experiment. While neglecting the PP contribution we obtain $P \sim
0.5$ (dashed line). In this figure, we show for comparison the
model results for CLAS energy $W=3.5 \mbox{GeV}$ by the dotted
line and for COMPASS energy $W=8 \mbox{GeV}$ by the dashed-dotted
curve. It can be seen that at COMPASS energies PP effects are
rather small for $\omega$. For the $\rho$ production PP effects in
$P$ asymmetry are shown in Fig. 5. They are small and the
asymmetry is close to unity.

\begin{figure}[h]
\begin{minipage}{17pc}
\includegraphics[width=17pc,height=6.0cm]{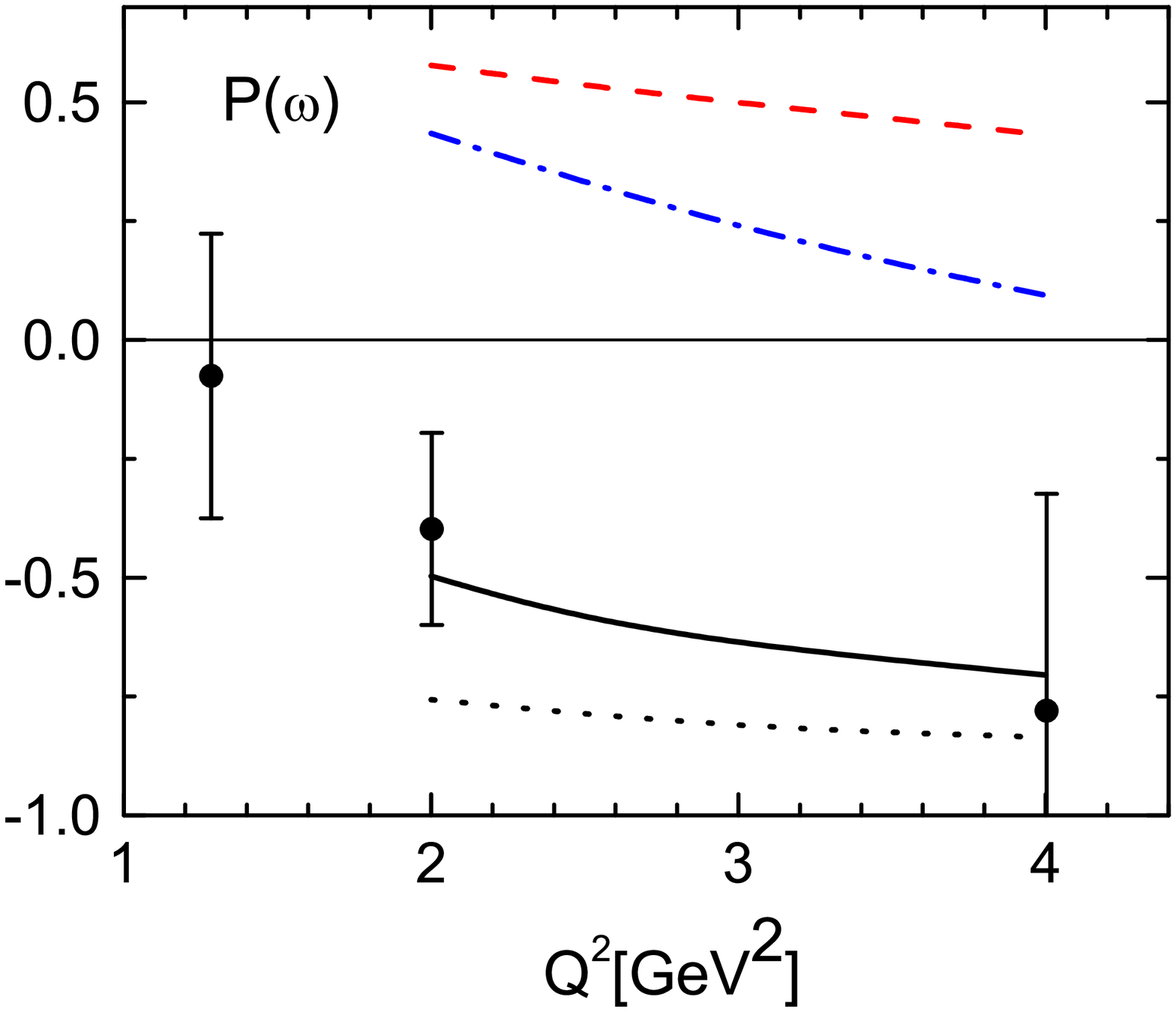}
\caption{\label{label2}
$P(\omega)$ at HERMES.  Solid line-with PP contribution, dashed
line -without PP. Dotted line-for $W=3.5 \mbox{GeV}$ (CLAS),
dashed-dotted line for $W=8 \mbox{GeV}$ (COMPASS)}
\end{minipage}\hspace{2pc}%
\begin{minipage}{17pc}
\includegraphics[width=17pc,height=6.1cm]{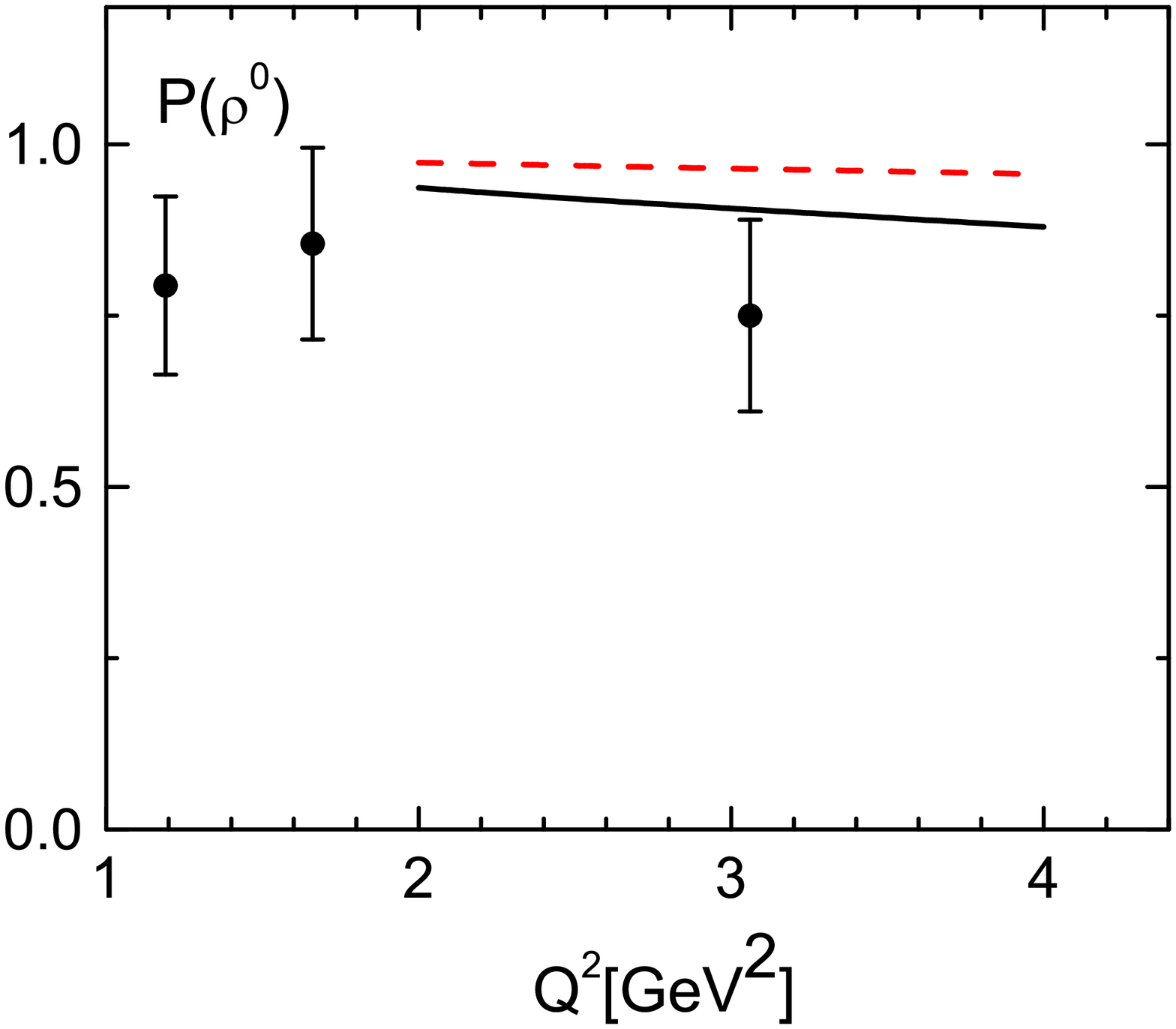}
\caption{\label{label2} $P(\rho^0)$ at HERMES. Solid line with
PP, dashed line -without PP.\\[8mm]}
\end{minipage}
\end{figure}

The SDME $r^{1}_{1-1}=- \mbox{Im}r^{2}_{1-1}$ shows the difference
of the natural and unnatural parity contributions
\begin{equation}
r^{1}_{1-1}  =\frac{d \sigma^N(\gamma^*_T \to V_T)-d
\sigma^U(\gamma^*_T \to V_T)}{2 d \sigma}.
\end{equation}
The results for this SDMEs for the $\omega$ production are shown
in Fig. 6. We see that the PP effects are very strong for
$\omega$. With PP contribution we obtain $r^{1}_{1-1} \sim -0.2$
and if we omit PP, $r^{1}_{1-1} \sim 0.2$. For the $\rho$ meson
production the PP contribution is small and the results are close
to the $\omega$ case without PP (see Fig.7). We find that HERMES
experimental results can not be explained without consideration of
the PP contributions.

\begin{figure}[h]
\begin{minipage}{17pc}
\includegraphics[width=17pc,height=6.2cm]{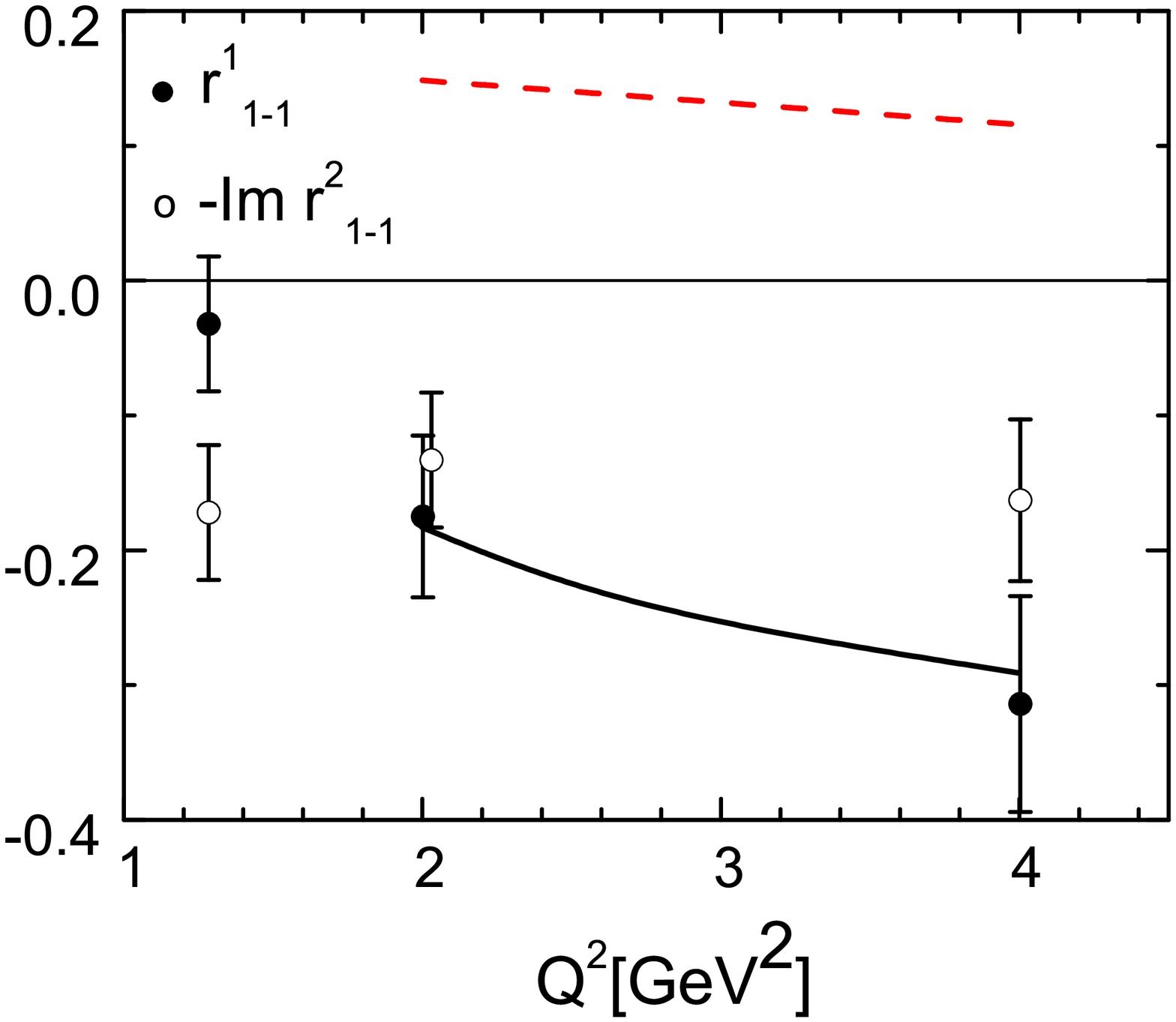}
\caption{\label{label2} SDMEs for $\omega$ production at HERMES. }
\end{minipage}\hspace{2pc}%
\begin{minipage}{17pc}
\includegraphics[width=17pc,height=6.1cm]{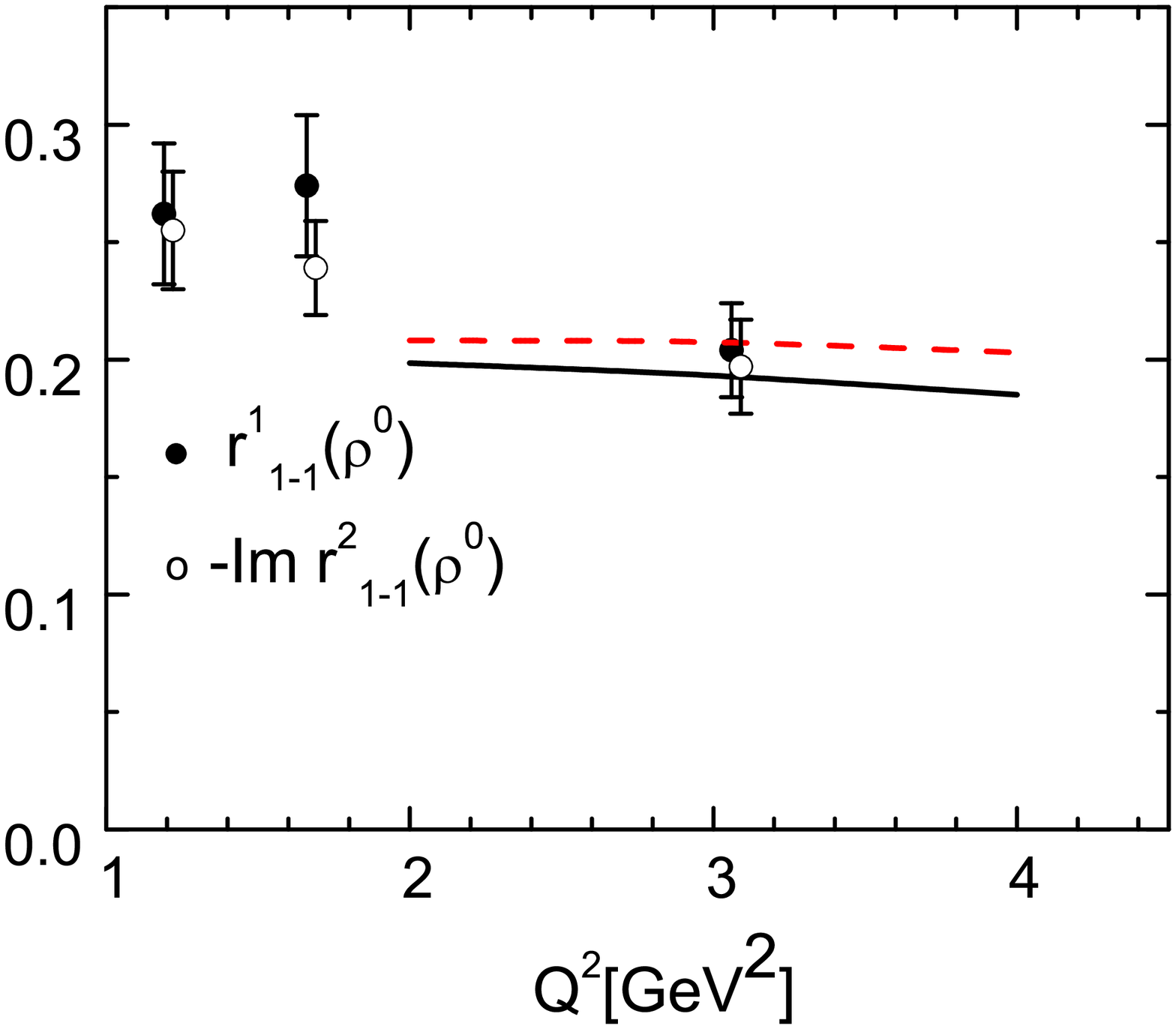}
\caption{\label{label2} SDMEs for
 $\rho$ production at HERMES.}
\end{minipage}
\vspace{3mm}

\centerline{  Solid line- with PP, dashed -without PP.}
\end{figure}

\begin{figure}[h]
\begin{minipage}{17pc}
\includegraphics[width=17pc,height=6.0cm]{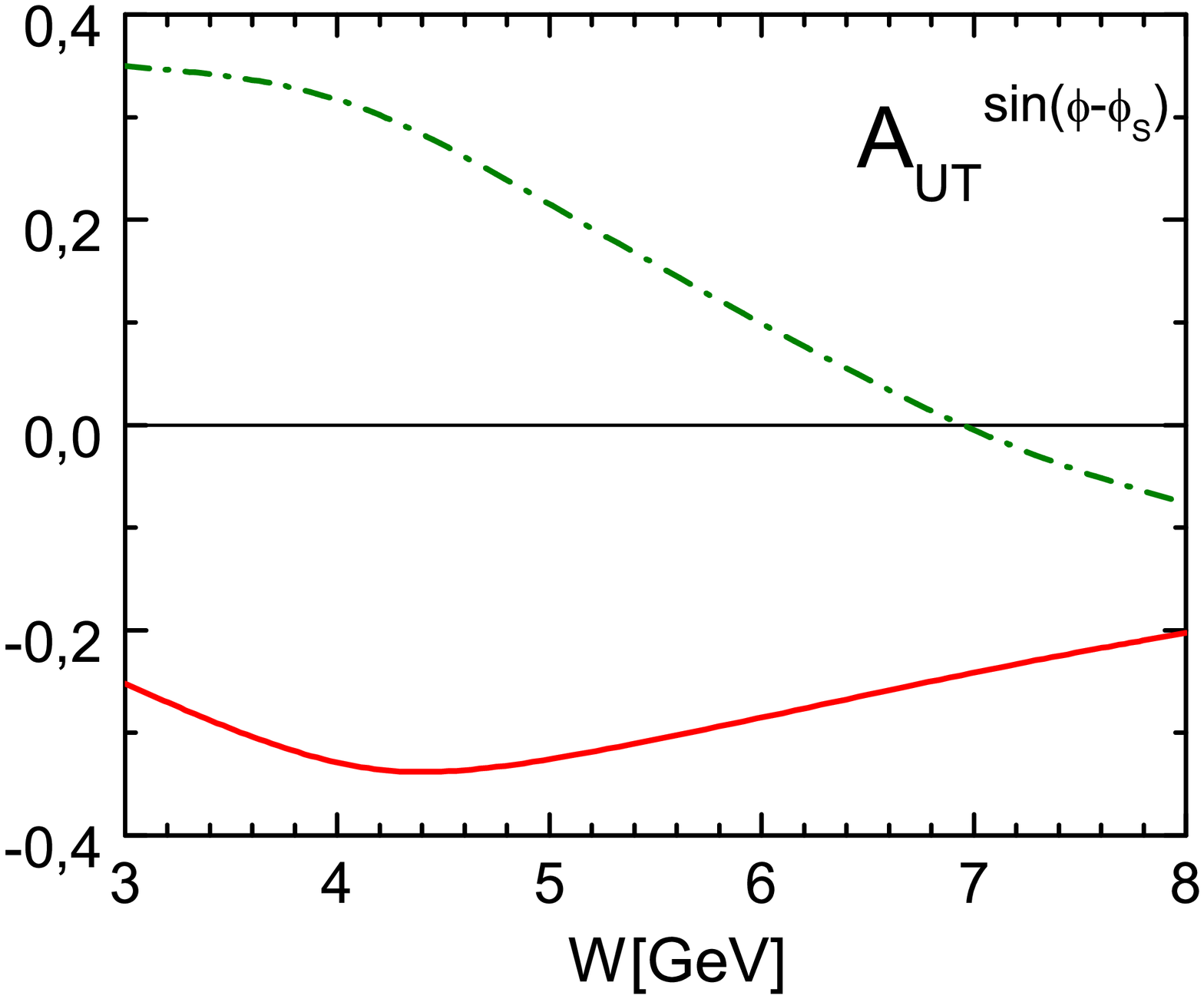} \caption{\label{label2}
 $\sin(\phi-\phi_s)$ modulation of $A_{UT}$ asymmetry.}
\end{minipage}\hspace{2pc}
\begin{minipage}{17pc}
\includegraphics[width=17pc,height=5.9cm]{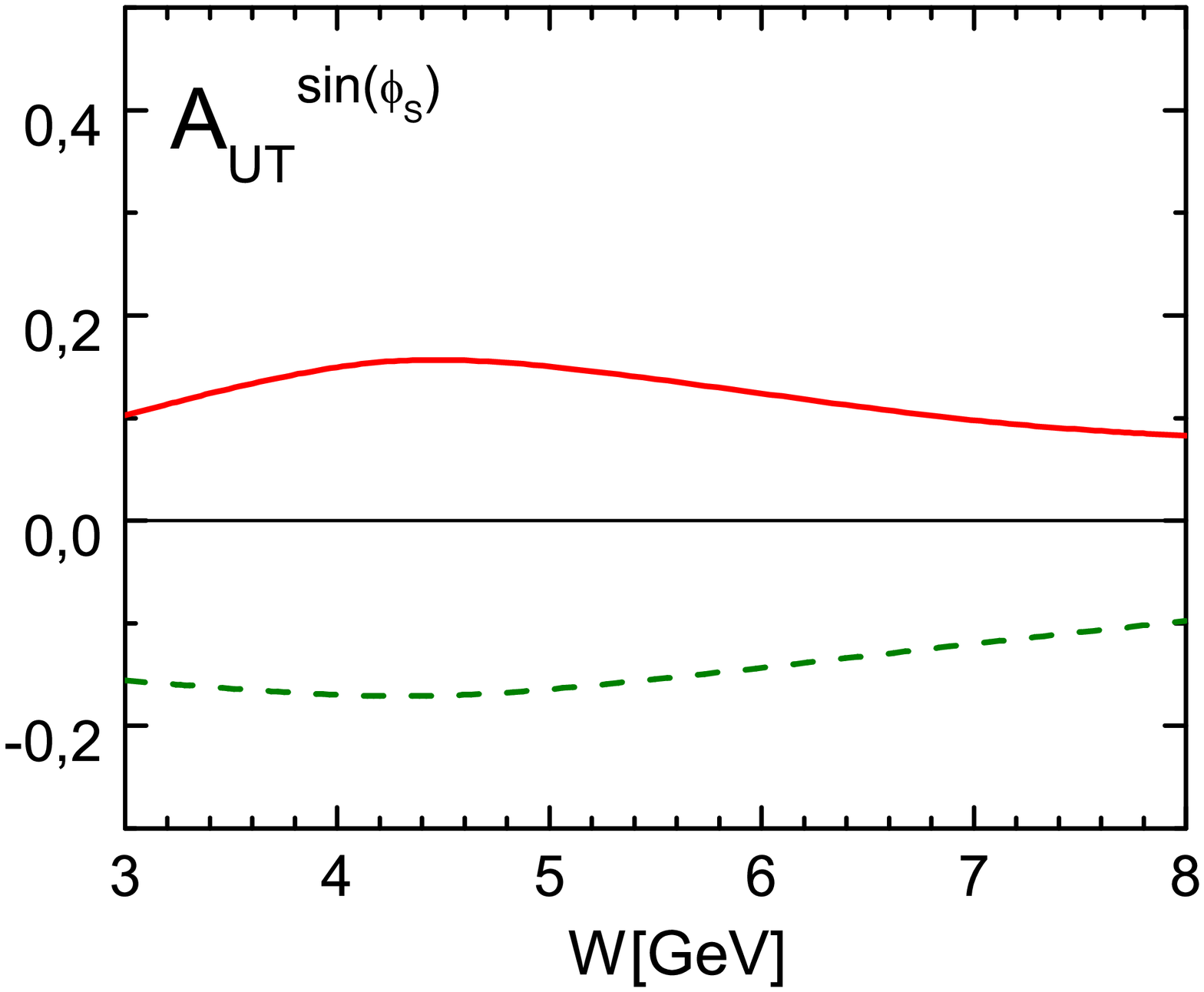}
\caption{\label{label2} $\sin(\phi_s)$ modulation of $A_{UT}$
asymmetry.}
\end{minipage}
\vspace{3mm}

\centerline{ Full line - positive, dashed- dotted line - negative
$\pi \omega$ transition form factor.}
\end{figure}
 \vspace{2mm}

  Pion pole effects  can be observed in spin
asymmetries. We find two not small interference terms with PP
\begin{equation}\label{asyup}
\mbox{Im} {M^{N*}_{++++}}\, M^{pole}_{+-++}\;\;\; \mbox{and}\;\;\;
\mbox{Im} {M^{U*}_{++++}}\, M^{pole}_{+-++}
\end{equation}
which contribute to spin asymmetries for longitudinal and
transverse beam and target polarizations. Note that the
$M^{U*}_{++++}$ amplitude in (\ref{asyup}) is determined by
$\tilde H$ effects. The PP contributions to asymmetries change the
sign with the transition $\pi V$ form factor.

Some examples for asymmetries in Figs. 8, 9 show substantially
different results for the positive and negative sign of the $\pi
\omega$ transition form factor. This means that generally, the
sign of this form factor might be determines experimentally.  New
HERMES data where 5 modulations of $A_{UT}$ asymmetry was analysed
\cite{hermesautomg} are probably consistent with the positive
transition $\pi \omega$ form factor. Unfortunately, the large
experimental uncertainties prevent a definite conclusion on the
sign of this transition form factor.

\section{Conclusion}
We  calculated the light meson leptoproduction  reactions  within
the handbag approach, where the amplitudes factorize into hard
subprocesses and  GPDs \cite{fact}.  The results  on the  cross
sections and various spin observables based on this approach are
in good agreement with data at HERMES, COMPASS and HERA energies
at high $Q^2$ \cite{gk06}.

We considered pion pole contribution to $\pi^+$ leptoproduction.
The PP effect in this case is determined by photon interaction
with a charged pion. It was shown that  PP  gave an essential
contribution to the leading twist longitudinal cross section
\cite{gk09}. Similar effects were found in the pion-induced
Drell-Yan process \cite{gkdy15}. The transversity $H_T$ and
$\bar{E}_T$ GPDs effects in the PM leptoproduction   were
investigated in \cite{gk09, gk11}. It was found that the
transversity contributions are essential in the PM leptoproduction
where they lead to large transverse cross sections which for
$\pi^0$ production at low $Q^2$ exceed essentially the leading
twist longitudinal cross section.

The pion pole effects in vector meson production have a different
nature. They are determined by the $\pi V$ transition form factor.
It is found that the PP contributions are essential in $\omega$
and rather small in $\rho^0$ production. Based on our approach we
found that the model results for $\omega$ SDMEs \cite{gk14} are in
good agreement with HERMES experimental results \cite{omega14}.
The large unnatural-parity effects observed by HERMES can be
explained by the substantial PP contributions there. The PP
effects in $\omega$ spin asymmetries are sensitive to the sign of
the $\pi \omega$ transition form factor \cite{gk14}. The
experimental test of these asymmetries by HERMES
\cite{hermesautomg} does not give a definite conclusion on the
sign of the $\pi \omega$ transition
form factor because of the large experimental uncertainties.\\

The work was supported in part by the Heisenberg-Landau program.

\end{document}